\documentstyle[emulateapj,apjfonts,flushrt]{article}
\makeatletter

\newlength{\bibhang}
\let\@internalcite\cite
\def\cite{\let\@citeleft(\let\@citeright)%
    \@ifstar{\citeyear}{\citefull}}
\def\citenp{\let\@citeleft\relax\let\@citeright\relax
    \@ifstar{\citeyear}{\citefull}}
\def\citefull{\def\astroncite##1##2{##1~##2}\@internalcite}
\def\citeyear{\def\astroncite##1##2{##2}\@internalcite}
\def\citeyp{\let\@citeleft\relax\let\@citeright\relax
    \@ifstar{\citeyearyp}{\citefullyp}}

\def\@citex[#1]#2{\if@filesw\immediate\write\@auxout{\string\citation{#2}}\fi
  \def\@citea{}\@cite{\@for\@citeb:=#2\do
    {\@citea\def\@citea{; }\@ifundefined
       {b@\@citeb}{{\bf ?}\@warning
       {Citation `\@citeb' on page \thepage \space undefined}}%
{\csname b@\@citeb\endcsname}}}{#1}}

\def\@cite#1#2{\@citeleft#1\if@tempswa , #2\fi\@citeright}
\def\@biblabel#1{}
\def\citefullyp{\def\astroncite##1##2{##1~(##2)}\@internalcite}

\makeatother

\newcommand{\PSbox}[3]{\mbox{\rule{0in}{#3}\includegraphics{#1}\hspace{#2}}}
\newcommand{\FigNum}[1]{\unitlength 1pt \begin{picture}(55,10)(-400,35) 
                        \put(0,0){Figure #1}
                        \end{picture}}

\newcommand{\persec}{\mbox{$\second^{-1}$}}

\newcommand{\percm}{\mbox{$\cm^{-2}$}}
\newcommand{\ppm}{\mbox{$\pm$}}
\newcommand{\cgsflux}{\erg\ \percm\ \persec}
\newcommand{\cgslum}{\erg\ \persec}

\newcommand{\lxlv}{\mbox{$L_x/L_{\rm bol}$}}
\newcommand{\ktr}{\mbox{$kT_{RS}$}}

\newcommand{\porb}{\mbox{$P_{\rm orb}$}}
\newcommand{\approxlt}{\mbox{$\lesssim$}}
\newcommand{\approxgt}{\mbox{$\gtrsim$}}
\def\etal{{et~al.}}

\def\x1608{{4U~1608$-$522}}
\def\xsix1608{{4U~1608$-$522}}
\def\x2129{{4U~2129+47}}
\def\cenx4{{Cen~X$-$4}}
\def\ufifteen{4U~1543$-$47}
\def\hseventeen{H~1705$-$25}
\def\nper{{GRO~J0422+32}}
\def\nmon{{A0620$-$00}}
\def\nmus{{GS~1124$-$68}}
\def\qzvul{{GS~2000+25}}
\def\v404cyg{{V404~Cyg}}
\def\nsco{{GRO~J1655$-$40}}

\def\fux{{$F_{\rm X, unabs.}$}}
\newcommand{\nh}{\mbox{$N_{\rm H}$}}
\newcommand{\nhtt}{\mbox{$N_{\rm H, 22}$}}
\def\aql{{Aql~X$-$1}}

\def\chisqrnu{\mbox{$\chi^2_\nu$}}

\newcommand{\ud}[2]{\mbox{$^{+ #1}_{- #2}$}}
\newcommand{\supp}[1]{\mbox{~~$^{#1}$}}
\newcommand{\ee}[1]{\mbox{$10^{#1}$}}
\newcommand{\tee}[1]{\mbox{$\times 10^{#1}$}}

\newcommand{\keV}{\mbox{$\rm\,keV$}}

\newcommand{\cm}{\mbox{$\rm\,cm$}}

\newcommand{\second}{\mbox{$\rm\,s$}}

\newcommand{\erg}{\mbox{$\rm\,erg$}}

\newcommand{\chandra}{{\em Chandra\/}}
\newcommand{\rosat}{{\em ROSAT\/}}
\newcommand{\asca}{{\em ASCA\/}}

\newcommand{\xmm}{{\em XMM\/}}

\newcommand{\av}{\mbox{$A_V$}}

\def\ma{m\AA}

\slugcomment{ApJ, Submitted Dec 21, 1999; Revised March 25,  2000;
Accepted May 5, 2000}
\begin{document}
\singlespace

\title{Coronal X-Ray Emission from the Stellar Companions to 
Transiently Accreting Black Holes}

\author{Lars Bildsten}
\affil{
Institute for Theoretical Physics and Department of Physics,
Kohn Hall, University of California, Santa Barbara, CA 93106\\
bildsten@itp.ucsb.edu}
\medskip

\author{ Robert E. Rutledge}
\affil{
Space Radiation Laboratory, California Institute of Technology, MS 220-47, Pasadena, CA 91125\\
rutledge@srl.caltech.edu}

\begin{abstract}

 Many neutron stars and black holes are in binaries where the mass
transfer rate onto the compact object is highly variable.  X-ray
observations of these transients in quiescence ($L_x< 10^{34} \ {\rm
erg \ s^{-1}}$) have found that the binaries harboring black holes are
much fainter than those that contain a neutron star. Narayan and
collaborators postulated that the faint X-ray emission from black hole
binaries was powered by an advection dominated accretion flow
(ADAF). The subsequent ADAF modeling requires that an appreciable
fraction of the constant Roche-lobe overflow into the outer disk
proceeds into the black hole during ``quiescence''. A robust and
nearly uniform quenching mechanism must then be hypothesized for the neutron
star binaries, as comparably large accretion rates would lead to
luminosities in excess of $10^{36} \ {\rm erg \ s^{-1}}$ in
quiescence.

We explore an alternative explanation for the quiescent X-ray emission
from the black hole systems: coronal emission from the rapidly
rotating optical companion. This is commonly observed and well studied
in other tidally locked binaries, such as the RS~CVn, Algol and By Dra
systems. We show that two of the three X-ray detected black hole
binaries (\nmon \ and \nsco) exhibit X-ray fluxes entirely consistent
with coronal emission. The X-ray spectra of these objects should be
best fit with thermal Raymond-Smith models rich in lines when coronal
emission predominates, a prediction that will be tested with {\it
Chandra} and {\it XMM-Newton} observations. One black hole system
(\v404cyg) is too X-ray bright to be explained as coronal emission,
and remains a candidate for ADAFs in quiescence. The quiescent X-ray
emission from all the neutron star binaries is far too bright for
coronal emission.  It might be that all SXT's have variable accretion
rates in quiescence and that the basal quiescent X-ray flux is set by
either coronal emission from the companion or -- when present -- by
thermal emission from the neutron star.

We have also searched for other indicators of coronal activity in
these companion stars. For example, we show that the lithium
abundances in the black hole systems are comparable to those in the
RS~CVns. Indeed, {\it both} the X-ray flux and lithium abundance from
the K star in the binary V471 Tau match that of \nmon. Though the
production mechanisms for lithium in active coronae is still under
debate, our work makes it clear that there is no longer a need for
mechanisms that involve the compact object.

\end{abstract}

\keywords{accretion, accretion disks --- binaries: close --- black
   hole physics --- stars: individual (\aql, \cenx4, \xsix1608,
   \x2129, \nmon, \nmus, \nsco, \v404cyg, GS~2000+25, \nper,
   \ufifteen, \hseventeen) --- X-rays: stars }

\section{Introduction}

 It is a mystery as to what powers the X-ray emission from transiently
accreting neutron star and black hole binaries (collectively referred
to as soft X-ray transients, SXTs) when they are in their faint,
quiescent state ($L_x< 10^{34} \ {\rm erg \ s^{-1}}$). The cause of
the bright outbursts is the sudden accretion of material that has
accumulated in the outer disk, just as in the dwarf novae systems (see
\citenp{king99} for a recent discussion). 
While the majority of the SXT's containing neutron stars have been
detected in quiescence, only three SXTs that contain black holes have
been detected: \nmon\ \cite{mcclintock95}, \v404cyg\
\cite{verbunt94,wagner94}, and \nsco\ \cite{hameury97}.  While there
is agreement that neutron stars (NS) are, on-average, brighter in
quiescence than the black-hole candidates (BHCs) 
\cite{barret96,narayan97a,asai98}, the sources of the quiescent
emission are still under debate.

Several possible mechanisms for the NS quiescent emission have been
introduced, including accretion \cite{jvp87,menou99} and
magnetospheric emission from a turned-on radio pulsar
\cite{campana98b}. Recent theoretical \cite{brown98} and observational
work \cite{rutledge99,rutledge00} successfully attributes much of the
quiescent luminosity of the transiently accreting NSs to thermal
emission from the NS surface. The repeated deposition of nuclear
energy deep in the crust (about 1 MeV per accreted baryon) during
outbursts keeps the core of these accreting neutron stars hotter than
their age would suggest \cite{brown98}. The luminosity of the thermal
emission is then fixed by the time-averaged accretion rate --
integrated over many outbursts -- to be $\approx 6\times 10^{32}$
\cgslum $(\langle \dot{M}\rangle/ \ee{-11} M_\odot \ {\rm yr^{-1}})$. 

  The only emission mechanism available to the quiescent black hole is
accretion. The puzzle of their emission mechanism began when
ROSAT/PSPC detected X-rays from \nmon\ at a level $L_x\approx 6\times
10^{30} \ {\rm erg \ s^{-1}}$ \cite{mcclintock95}.  McClintock et
al. (1995) made it clear that this X-ray emission could not be due to
a steady-state accretion disk around the black hole, as if so, there
would be a production of optical and UV photons from the outer parts
of the disk which would far exceed that observed. This puzzle can be
solved with an advection-dominated accretion flow (ADAF), making it a
commonly discussed energy source for the BHC quiescent X-ray emission
\cite{narayan97a,narayan97b,yi97,quataert99}.  In this case, the
X-rays are produced via Compton up-scattering of the optical/UV
synchrotron emission from the flow close to the black hole. These
models much more successfully handle the ratio of optical/UV emission
to X-ray emission, which is critical in evaluating their success (\S
\ref{sec:adaf}). These models require large accretion rates in
quiescence; fully a third of the mass transfer rate in the binary (see
\citenp{meyerhofmeister99} for a discussion).

 In this paper, we investigate the alternative hypothesis that 
some or all of the quiescent X-ray emission from the BHCs originates
from the active corona of the rapidly rotating and convective stellar
companion. Coronal X-ray emission from the companion star has not
been examined in any detail previously.  McClintock et al. (1995)
concluded that the X-ray luminosity of coronally active stars was too
low to explain BH SXT emission ($L_x<$\ee{30} \cgslum), referring to a
comparison between the X-ray luminosity of CVs and unevolved K dwarfs
in the Pleiades made by Eracleous \etal \cite*{eracleous91}.
However, the more analogous systems are tidally locked (and
slightly evolved) stars in tight binaries, such as the RS~CVn and
Algol systems. These have X-ray luminosities from coronal activity
that can reach \ee{32} \cgslum \cite{dempsey93a}. 
The analogy between quiescent BHCs and the
RS CVn binaries was further noted by Verbunt (1996)\nocite{verbunt96},
and we confirm his initial findings about the X-ray emission.

In \S~\ref{sec:rscvn}, we briefly discuss coronal X-ray production in
other tidally locked binaries, making it clear what level of X-ray
emission is possible from the SXT companion stars. We find that the
X-ray emission from two (\nmon \ and \nsco) of the three detected BHCs
is consistent with coronal emission, while one BHC (\v404cyg) and all
NS systems are too bright to be explained with coronal emission from
the companion. The implications this has for future X-ray observations
are discussed in \S \ref{sec:futxray}. Possible radio and/or optical
confirmation of the coronal hypothesis is discussed in \S
\ref{sec:implications}.  When first noting the prominent Li absorption
line in the SXT \v404cyg, Wallerstein \cite*{wallerstein92} considered
the hypothesis that Li was overabundant as in an RS CVn binary. Though
he discarded it because he estimated the Li to be too abundant, we
feel this was premature.  Our investigations into coronal indicators
find that the lithium detected in many of the SXT companions is likely
from coronal activity, thus eliminating the need for exotic mechanism
having to do with the compact object. We conclude in
\S~\ref{sec:conclude} and mention a few tests for coronal X-ray
emission in these systems. We also outline the implications our work
has for the general picture of accretion in quiescence in the SXTs.

\section{Stellar Coronal X-ray Activity and Comparisons to SXTs}
\label{sec:rscvn}

The level of coronal activity for stars with outer convective zones
strongly depends on the stellar rotation rate (see
\citenp{noyes84}). The simplest dynamo picture requires rotation and a
radiative layer underneath the convective zone. This naturally occurs
in late-type main sequence stars (spectral types F through M) where
the accumulated X-ray observations point to the ratio of soft X-ray to
bolometric luminosities ($L_x/L_{\rm bol}$) saturating at a
level of $\approx 10^{-3}$ \cite{vilhu87} toward short rotation
periods (\approxlt 1 day) \cite{singh99}. This saturation level
decreases for rotation periods longer than a few days  and appears to
be independent of stellar age \cite{singh99}. 

The inevitable slowing of isolated stars with age means that high
levels of activity can best be maintained for long times via tidal
locking in a tight binary. The common systems that result are the RS
CVn, Algol and By Dra type binaries; many of which are prevalent X-ray
sources in old open clusters, such as M67 \cite{belloni98}.  The X-ray
properties of RS CVns were surveyed by Dempsey \etal\
\cite*{dempsey93a}, who found 112 such objects in the \rosat\ All-Sky
Survey. They found that most rapidly rotating (less than ten days)
late type dwarfs exhibit $L_x/L_{\rm bol}\approx 10^{-4}-10^{-3}$,
with some systems exceeding the $10^{-3}$ level that Vilhu and Walter
\cite*{vilhu87} had denoted previously as a possible ``saturation''
limit. Dempsey et al. \cite*{dempsey97} found much the same X-ray
properties for the BY Draconis dwarf-type binaries.

The companion stars in the SXTs are also rapidly rotating through
tidal locking and have rotation rates and spectral types similar to
these active systems \cite{verbunt96}.  This amplifies the need to
carefully assess the likelihood that a significant part of the
observed soft X-ray emission in these binaries is from coronal
activity.  We start by making the simplest comparison between the SXT
companions and the active binaries, which is the ratio of the X-ray to
bolometric flux, \lxlv. This {\it distant-independent} quantity
immediately tells us which SXTs could possibly have substantial X-ray
emission from active coronae.

\subsection{Measuring \lxlv\ for the SXTs}
\label{sec:calc}

The SXTs we analyze are the optically identified, spectroscopically
typed SXTs with low quiescent luminosities (\approxlt \ee{33} \cgslum;
\citenp{chen97,menou99}).  This gives us eight BHCs: \nper, \nmon,
\nmus, \nsco, \ufifteen, \hseventeen, \qzvul \ and \v404cyg; and four
NSs: \aql, \cenx4, \xsix1608, and \x2129.  

Here, we give the details of measuring \lxlv\ for the SXTs.  To obtain
X-ray fluxes in the \rosat/PSPC passband (0.4-2.4 keV) for comparison
with coronal sources, we re-fit X-ray data from our previous studies
of these objects \cite{rutledge99,rutledge00}, and converted the
upper-limit fluxes for those we had not previously analyzed
(specifically, \ufifteen \ and \hseventeen) using
W3PIMMS\footnote{http://heasarc.gsfc.nasa.gov}.  For the fits, we
assume all the X-ray emission is from the stellar corona, and use a
Raymond-Smith plasma plus galactic absorption column, as implemented
in XSPEC v10.0 \cite{xspec}, with solar-abundance, and holding the
column density \nhtt(=\nh/[\ee{22} \percm]) fixed at the values we
previously adopted for each source, except where noted.  To obtain
2$\sigma$ upper-limits, we assume \ktr=1.0 keV.

The BHC data we analyse here are of low signal to noise.  For example,
\nmon \ has only 45 +/- 9 counts.  Binned into 3 PHA spectral bins,
these data can be adequately fit with nearly any model (power-law,
blackbody, bremmstrahlung).  It is our present goal to only interpret
the X-ray flux as from a stellar corona, which motivates our selection
and application of a Raymond-Smith plasma model. The current data do
not allow us to test the hypothesis based on the spectral fits.  These
data have been previously analysed in a wide range of models, and we
refer the interested reader of the results of such analyses to those
studies \cite[and references therein]{rutledge00}.

After measuring the unabsorbed X-ray flux \fux\ (0.4-2.4 keV), we also
find the bolometric luminosity of the stellar companion 
$F_{\rm bol} = {10^{-0.4(V_q +  \delta V_{\rm  exc} + 11.51 +
B.C. - A_V)}}$ \cgsflux \cite[p. 102]{zombeck}, where $V_q$ is the
quiescent magnitude, $\delta V_{\rm exc}$ is a correction
due to the non-stellar continuum (see \S \ref{sec:adaf} 
for a discussion), B.C. is the bolometric
correction for the measured spectral type, and \av\ is the reddening.

To estimate $\delta V_{\rm exc}$ we examined all previous measurements
of the fraction ($f$) of the optical continuum that is not from the
stellar companion (see Table~\ref{tab:f}). These measurements were taken
in different passbands, from different objects, often using different
techniques and at different times, some with and others without error
estimation. The vast majority are $f\approxlt$50\%.  Four
measurements of the $\sim$30 we list are above this and we therefore
consider a $>50$\% flux contribution to be exceptional (the data of
\citenp{harlaftis96} is the same as \citenp{filippenko97}).  Note
especially the values by Orosz \etal \cite*{orosz96} for \nmus\ and
\cite*{orosz97} for \nsco, which indicate a time-variable $f$.  While
it would be highly preferable to apply individual --- and more
precise --- estimates of this contribution for the individual sources,
such contributions must be measured simultaneously with the
photometric measurement, due to the observed time-variability of $f$,
and these are generally not available.  We adopt $\delta V_{\rm
exc}=0.4\pm0.4m$, which includes a conservative uncertainty for this
correction. Aside from uncertainty in disk contribution we estimate
the uncertainty in $F_{\rm bol}$ to be dominated by uncertainty in the
B.C. and $A_v$.  Our (conservative) uncertainties in the B.C. are
found from the effect of the uncertainty of the companion's spectral
type on the assumed B.C.  The uncertainties in $V_q$ are small
compared to these, so we neglect them.

We use the following conversions: $A_V=3.1 \; E(B-V)$,
\cite{schild77}; \nh=0.179 $A_V 10^{22}$cm$^{-2}$ \cite{predehl95}.
We now describe the details of the calculation as it differs from this
description, as applied to each source, and the statistical quality of
the best-fit. Table~\ref{tab:xrt} shows the derived \fux\ and $F_{\rm bol}$ values
as well as the information used to derive them.

{\it Aql X-1}. As in our previous work \cite{rutledge99}, we used the
two ROSAT/PSPC observations and the ASCA observation in a
simultaneous fit, imposing an identical spectrum, but permitting the
three values of the column density to vary independently (if we
require all three column densities to be equal, the fit is
statistically unacceptable, as previously).  The spectral fit was only
marginally acceptable (\chisqrnu=1.38/78 degrees of freedom=dof;
prob=0.015), and \ktr=1.38\ppm0.05\keV.

{\it Cen~X-4}.  We let the column density vary.  While the best-fit
spectrum was statistically unacceptable (\chisqrnu=1.82 for 90 dof;
prob=3\ee{-6}), we use this to estimate the \fux (we estimate the flux
uncertainty at 10\%, although the spectral uncertainty is greater). 


{\it \xsix1608}.  We alternately adopted the two values of \nhtt=0.8
and 1.5 (cf. \citenp{rutledge99,wachter97}).  We thus find
\ktr=1.2\ppm0.1 keV (0.6\ppm0.04 keV) and \fux=1.6\tee{-12}
(1.3\tee{-11}) \cgsflux for \chisqrnu=1.52 (1.34; 61 dof) with
probabilities prob=0.004 (prob=0.03) for the two listed \nh\ values,
respectively.  The uncertainties in the flux values were
\ud{40\%}{20\%}.  No orbital period and only loose constraints on the
spectral type of the companion QX Nor in quiescence (luminosity class
IV-V, spectral class G0-K0, assuming $d$\approxgt3.0 kpc) have been
obtained for this source \cite{wachter97}.  We adopt the quiescent
$J=18$ of Wachter \cite*{wachter97}, $(V-J)_0=$1.05--1.43
\cite[p. 68]{zombeck}, and the values $A_J=1.6-2.5$ (for low and high
values of \nh). From this we obtain $V_q$=19.25\ppm0.2

{\it 4U 2129+47}.  We re-fit our previous PSPC spectrum with
\nhtt=0.22 \cite{deutsch96}, obtaining \ktr=0.82\ppm0.06 \keV
(\chisqrnu=0.3/1 dof; prob=0.6). For the HRI data, we held \nhtt=0.22
and \ktr=0.82 keV fixed.  We obtained \fux=1.2\tee{-13} and
2.1\tee{-13} \cgsflux\ for the PSPC and HRI observations,
respectively.

{\it \nmon}.  We obtained a best fit spectrum with
\ktr=0.6\ppm0.4\keV, with an acceptable fit of \chisqrnu=2.8/1 dof
(prob=9\%).


{\it \ufifteen}.  We converted the X-ray flux 5$\sigma$ upper-limit
\cite{orosz98}, using our adopted \nh=0.5 and \ktr=1.0 keV, to estimate
a 2$\sigma$ upper-limit.

{\it GRO J1655-40}.  The fit to \asca\ data was acceptable
(\chisqrnu=1.2/16 dof; prob=0.26).  With both the temperature and
normalization allowed to vary, the data provides only a lower limit
for temperature of \ktr$\geq$ 4.1\keV (90\%), with an upper-limit in
excess of 64\keV. At the best-fit, the uncertainty in flux (holding
the temperature fixed) is 10\%; across the range of temperatures
4.1-64\keV, the best fit flux ranges 30\%. The flux measurement
uncertainty then is dominated by uncertainty in the spectrum.  We
quote the best fit X-ray flux (6\ppm1.2)\tee{-14}
\cgsflux, adopting a 20\% uncertainty.

{\it \hseventeen}. We used the 4$\sigma$ HRI countrate upper-limit
(0.006 c/s; \citenp{verbunt94}), \nhtt=0.27, and \ktr=1.0 keV to
estimate a 2$\sigma$ upper limit.

{\it GS 2000+25}.  We estimate the $V_{\rm q}$ from $R$=20.8
\cite{chev90}, and $(V-R)_0$=0.99 of a K5V star
\cite{zombeck}. \label{sec:2000}

{\it \v404cyg}.  These data are such that we only find a lower limit
of \ktr$>$1.5 \keV (90\% confidence), where the flux is 78\tee{-14}
\cgsflux.  At \ktr=64\keV, the flux is 96\tee{-14} \cgsflux.  The
uncertainty in flux (holding temperature fixed) is 5\%; the X-ray flux
uncertainty is then dominated by spectral uncertainty, where it is
20\%. We adopt a flux value of (85\ppm20)\tee{-14} \cgsflux.  

The consistency of these spectral fits with a Raymond-Smith plasma is
not, in itself, supporting evidence for a coronal origin of the
observed X-ray flux, as the data S/N are such that nearly any spectrum
(black-body, power-law) would be consistent with them (see
\cite{rutledge00} and references therein).  Rather, we applied
Raymond-Smith spectra only to extract X-ray fluxes from the data, to
produce the ratio $L_x/L_{bol}$, assuming that the X-rays are coronal
in origin.  We cannot rule out, on the basis of these X-ray spectra,
that some other emission mechanism is responsible for the X-ray flux.

\subsection{Comparison of \lxlv\ between SXTs and Active Binaries}

The measured ratios (\lxlv) are displayed in Figure~\ref{fig:lxlbol}
for the SXTs, as well as the RS~CVNs \cite{dempsey93a}, as a function
of the binary orbital period.  We also include the best-fit line
relating \lxlv\ and rotation period for isolated late-type dwarfs
\cite{singh99}.  The \lxlv\ ratios for the SXTs are uncertain by up to
a factor of two, depending on the source.  Two of the three X-ray
detected BHCs (\nmon \ and \nsco, the exception being \v404cyg) have
\lxlv\ ratios common to the RS~CVns (\approxlt \ee{-3}).  The four
detected NSs are 10-1000 times above this value, clearly inconsistent
with coronal X-ray emission. Five other BHCs (\nper, \ufifteen,
\hseventeen, \nmon, \qzvul) have upper limits (some high, others less
so) which are also consistent with RS CVn-like X-ray emission. {\it We
conclude that the quiescent X-ray emission of some BHCs can be
dominated by the coronal X-ray emission of the companion. This 
is not the case for the quiescent emission of NSs.}

Among the BHCs, \v404cyg is clearly an outlier in \lxlv.  The
quiescent source is also spectrally harder than other BHCs and NSs
\cite{rutledge00}.  It is unclear if this can be associated with its
highly variable outburst X-ray intensity profile, radio intensity
profile and polarization, absorption column, and Fe line -- all
dramatically different from those observed from other transient BHCs
\cite{kitamoto89,han92,timo96}.

  The dwarf novae (DN) are another class of transiently accreting
binaries, where white dwarfs undergo large accretion events after long
periods of quiescence. They were found by {\it Einstein} to be X-ray
sources (\ee{30}--\ee{32} \cgslum) in quiescence
\cite{cordova84,patterson85}; raising much the same issues as we have
discussed for the SXTs. However, in this case, Eracleous \etal\
\cite*{eracleous91} clearly showed that the X-ray luminosities of the
late-type rapidly rotating main sequence stars (that are the
companions to these white dwarfs; \citenp{smithdhillon98}) are too low
(\approxlt \ee{30} \cgslum; \citenp{cruddace84,fleming89}) to explain
the quiescent X-ray luminosities. Moreover, these CVs have
\lxlv$\sim$\ee{-2}--10 \cite{richman96}, well above the \ee{-3}
saturation upper-limit observed in chromospherically active systems.
This indicates that the majority of the measured X-ray emission is
from accretion in quiescence. The typical quiescent accretion rate is
a few percent of that transferred in the binary, and the X-rays should
originate from the boundary layer near the white dwarf
\cite{patterson85}. This has been confirmed via X-ray eclipse
observations with ROSAT of three short orbital period DN in
quiescence. These are HT Cas at $P_o=1.768$h \cite{mukai97}, Z Cha at
$P_o=1.788$h \cite{vanteeseling97} and OY Car at $P_o=1.515$h
\cite{pratt99}. All of these showed that the measureable X-ray
emission was completely eclipsed when the white dwarf was behind the
companion. No similar observations have yet been done for wider
orbital period systems analogous to \nmon. 

\section{Implications for Future X-Ray Observations of the BHCs}
\label{sec:futxray} 

We have shown that the quiescent X-ray emission from all but one BHC
might be due to coronal emission from the companion star. Figure 2
shows the quiescent X-ray flux versus the bolometric flux of the
companion star for the BHCs.  The horizontal line displays the flux
that can be reached in a 50 ksec \chandra\ observation with ACIS-S
(assuming 1 count=5\tee{-12} ergs \percm, with 10 counts required for
detection). This highlights what progress can be made.  If most of
the X-ray emission from the quiescent BHCs is of a coronal origin, it
is unlikely that GRO J0422+32, \hseventeen \ or GS 1124-68 will be
detected.  However, GS~2000+25 stands out as a potential
\chandra\ detection. The stellar companion in GS 2000+25 is a K3-K6
dwarf tidally locked in a 0.344 day orbit
\cite{chev93,filippenko95b,harlaftis97}, much like A0620-00, and we
expect coronal emission about at the \chandra\ detection limit.

 Unlike all of the other BHC companion stars (which are later than mid
F type, see Table~\ref{tab:xrt}), the stellar companion in 4U~1543-47
is an early type star of spectral type A2 V \cite{orosz98}. Hence,
though it appears to be a detectable source with {\it Chandra}, we
don't expect it to be detected as a coronal source. 
Our earlier
discussion of coronal activity was independent of spectral type for
mid to late-type stars. The story is different for A stars, where soft
X-ray emission is at least two orders of magnitude weaker than in
comparably rapid rotating late type stars \cite{simon95}. The likely
reason is the declining strength (or even absence) of an outer
convective zone in stars this hot. The brightest X-ray emission
detected from an early A star is the rapid rotator HR 5037 (which has
the same spectral type and rotation rate of the companion to
4U~1543-47) at $L_x/L_{\rm bol}\approx 1.5\times 10^{-5}$
\cite{simon95}. Most other early A stars have upper limits at levels a
factor of ten lower.  Hence, we expect that no coronal emission will
be detected from this companion, making it a ``clean'' system for
studying alternate energy sources for quiescent X-ray emission, such
as the ADAFs.  

   We also examined the prospects for detecting (with a 50 ksec
\chandra\ observation) coronal X-rays from all BHCs listed in Chen
\etal \cite*{chen97}, or discovered since that compilation. 
We require a spectral type, quiescent $V$ magnitude and
$A_V$ in order to estimate the bolometric flux.  Other than those we
discuss elsewhere in this paper (for which there are stringent
quiescent X-ray flux limits or measurements, see Fig.~\ref{fig:lars}),
only N Vel 1993 (=1009$-$45) has such values reliably measured
($R$=21.2\ppm0.2, K7-M0V, \citenp{filippenko99}; and
$E(B-V)$=0.2\ppm0.05, \cite{dellavalle97}).  For $(V-R)_0=1.15$,
B.C.=$-0.9$ \cite{zombeck}, $A_R$=2.3$E(B-V)$ \cite{fitzpatrick99},
and assuming \lxlv=\ee{-3}, we find the X-ray flux due to the corona
should be \ee{-16} \cgsflux, well below the \chandra\ 50 ksec detection
limit.

 With regards to XTE J1550-564, Jain \etal\ \cite*{jain99} find a
$B\approx22$ source at the position of the X-ray object in quiescence,
speculating it to be a type G V object, with $A_V=5.0$ (an upper-limit
based on radio measurements). For a G5V source, $(B-V)_0=0.66$
\cite{zombeck}, and B.C.$=-0.1$, producing $F_{\rm bol}\sim8\times
10^{-12}$ \cgsflux. For \lxlv$\sim$\ee{-3}, this implies
\fux$\sim$8\tee{-15} \cgsflux. Alternatively, for a K0V star, with
B.C.$=-0.19$ and $(B-V)_0=0.82$, \fux=10\tee{-15}. These unabsorbed
fluxes (with \nhtt=0.9) can be detected with \chandra\ or \xmm.
However, Sobczak \etal\ \cite*{sobczak99} found through X-ray
spectroscopy a value of \nhtt=1.7-2.2, which for interstellar
absorption (\nhtt=0.179 $A_V$) implies $A_V\sim$ 9.5-12.2. This
implies quiescent bolometric and coronal X-ray luminosities a factor
of 100 higher than these estimates. Thus, the uncertainty in the
expected coronal X-ray luminosity is dominated by the uncertain $A_V$.

\section{Indications of Coronal Activity at Other Wavelengths}
\label{sec:implications} 

 Though our work was motivated by the X-ray measurements, there are
numerous other repercussions if the SXT companions are coronally
active. Out of all the things we discuss here, perhaps the most
important is our realization that the lithium seen in these systems is
possibly  from coronal activity.

\subsection{Lithium Absorption Lines in the SXT Companions} 

  The $\lambda6708\AA$ Li line is an absorption line that originates
in the stellar atmosphere of the companion. It has has been found to
be anomalously strong (when compared to field stars of the same
spectral and luminosity class) in \v404cyg\ (equivalent width
EW=290\ppm50 \ma ; \citenp{martin92,martin94a}), \nmon\ (160\ppm30 \ma
; \citenp{marsh94}) \cenx4\ (480\ppm65 \ma; \citenp{martin94a});
\qzvul\ (270\ppm40 \ma ; \citenp{filippenko95b,harlaftis96}), and
\nmus (420 \ppm 60 \ma ; \citenp{martin96}).  Upper-limits on Li in
\aql\ ($<$300 \ma, \citenp{garcia99}) and \nper\ ($\le480$ \ma,
\citenp{martin96}) are consistent with other SXT Li measurements.
Following these obsevations, a number of Li production mechanisms
which require the presence of a compact object were suggested
\cite{martin94b,yi97,guessoum99}; these have little {\it a priori}
ability to predict the level of lithium abundance.

 Lithium is also observed in RS~CVns and other active stars with
abundances above that from field stars
(\citenp{pallavicini92,randich93,fernandez93,randich94}). The origin
of the Li overabundance in the active systems remains unclear;
however, there is a claimed statistical correlation --- with a broad
dispersion -- between Li abundance and Mg {\sc II}, which is used as a
coronal activity indicator \cite{zboril97,zboril98}. If the SXT
companions are chromospherically active, then their Li overabundance
may be due in part to the same mechanisms -- whatever these may be --
which produce them in chromospherically active stellar systems.  In
Fig.~\ref{fig:li}, we show the distribution of detected Li abundances
in chromospherically active systems \cite{pallavicini92} and five SXTs
\cite{martin94a,martin96}. Though slightly overabundant relative to
the RS~CVns, we see no need for production mechanisms special to the
compact object.

 It is also of interest to compare A~0620-00 with the pre-CV binary V
471 Tauri, which contains a white dwarf and a K2V star in a 12.5 hour
orbit. Wheatley \cite*{wheatley98} recently showed that the hard X-ray
emission from this system is not eclipsed by the K star, demonstrating
clearly that the emission is coronal from the K star at a level of
$L_x/L_{\rm bol}\approx 10^{-3}$. Martin \etal\ \cite*{martin97} had
earlier found the lithium absorption line present with an equivalent
width of 299 \ma, and derived an abundance $N(\rm Li)\approx
2.3$. These parameters are all very similar to the K star companion to
A~0620-00.  The simplest hypothesis for both the X-rays and lithium is
coronal activity.

\subsection{Radio Emission from Stellar-Coronal Sources}

Radio emission (4.9-8.5 GHz) has been compared with the soft X-ray
flux measured with ({\it Einstein} and \rosat) from stellar coronae of
dM(e), dK(e), By Dra-type, RS CVns, and Algols \cite{benz94} in which
it was found that the X-ray and radio luminosities are correlated over
a range of \ee{8} in X-ray luminosity (as high as $10^{32}$ \cgslum)
and a range of \ee{10} of radio luminosity, represented by
\cite{guedel93,benz94}:

\begin{equation}
\label{eq:radio}
F_{\rm  Radio} = 0.32 \left( F_x \over 10^{-14} \;  {\rm erg \;  cm^{-2}
\; s^{-1}}\right) \frac{10^{\pm0.5}}{\kappa} \, \mu{\rm Jy}, 
\end{equation}

\noindent where $\kappa$ is $\sim0.17-1.0$ ($\kappa$=1.0 for dMe, dK,
BY Dra and RS CVNs with two sub-giants, and $\kappa=$0.17 for the
higher X-ray luminosity, \approxgt\ee{30}\cgslum classical RS CVNs,
Algols, FK Com and post T-Tauri stars).  

For the quiescent X-ray fluxes of the detected BHCs in Table~\ref{tab:xrt}
(between 1.2--95\tee{-14} \cgsflux) the corresponding radio fluxes are
between 0.4--30 $\mu$Jy, which are below reported upper limits for
these sources: ($<$300 $\mu$Jy at 4.8 GHz for \nmon, $<$300 $\mu$Jy at
4.8 GHz for \nper, \citenp{geldzahler87}; $<$500 $\mu$Jy at 4.8 GHz
for GRO J1655-40, \citenp{hjellmingiauc96}).  This expected coronal
flux density level is at or below the detection limits of present
instrumentation, representing an observational challenge.  On the
other hand the factor of 10 dispersion in the X-ray/radio flux ratio
may provide a fortuitous detection.  If \v404cyg\ X-ray emission were
coronal, we would expect it to be detected with $F_{\rm
radio}\sim30-200\mu$Jy while at the reported X-ray flux.

\subsection{Optical and Ultraviolet Emission Lines} 

There are a few studied optical and UV emission lines which
are found in spectra of coronally active systems, including Mg {\sc
II}, H$\alpha$,  and C
{\sc IV}.   Here, we compare observations of these lines in SXTs to
the same in the coronally active systems. 

The Mg {\sc II} H and K emission lines are indicative of chromospheric
activity in the RS CVn binaries \cite{rucinski85,fernandez86} and
have been detected in A0620-00 \cite{mcclintock95,mcclintock99}, Cen
X-4 \cite{mcclintock99} and GRO J0422+32 \cite{hynes99}. In A0620-00,
the line shape was broad and single peaked, suggestive of a stellar
origin from the star, rather than from a disk (where the lines are
double peaked). The observed fluxes are $F({\rm Mg II})>10^{-3} F_{\rm
Bol}$ in Cen X-4 and A0620-00 and $F({\rm Mg {\sc II}})\approx 10^{-2}
F_{\rm Bol}$ in GRO J0422+32. These flux ratios are more than a factor
of ten higher than is observed in even the most active binaries, where
at maximum rotation rates, $F({\rm Mg {\sc II}})/F_{\rm Bol}\approx
2\times 10^{-4}$ (see \citenp{rucinski85,fernandez86}). Even when the
X-ray flux ratio is approaching $10^{-3}$ in the active binaries (as
we see in some of the BHC's), the flux ratio in the Mg {\sc II} lines
are still $<3\times 10^{-4}$.  As an example, \citefullyp{dempsey96}
found Mg {\sc II} fluxes about a factor of ten below that of the
X-rays from the KI IV companion in the active binary V711 Tauri.  The
only possible way to explain the Mg {\sc II} line emission from the
BHC's as coronal is to attribute it to higher abundances of this
$\alpha$ element, as recently reported by \citefullyp{israelian99} in
GRO J1655-40. It seems more likely that the Mg {\sc II} emission
originates in the accretion disk. 

 Another suggestive indication of coronal activity exists for Nova Oph
1977 (H 1705-25).  Harlaftis \etal\ \cite*{harlaftis97} reported on
their Doppler mapping of the H$\alpha$ line emission, which allows for
a secure identification of this emission with the outer accretion disk
and the ``splash point''. In addition, they also reported evidence for
H$\alpha$ emission from the companion star itself. This might be
indicative of coronal activity. However, they did not report the flux
for this component and so we cannot compare it to the typical levels
seen in other interacting binaries ($L_{\rm H_\alpha}/L_{\rm bol}\sim
3-16 \tee{-5}$; \citenp{barden85,montes95}).  From our estimate of
$F_{\rm bol}$, and assuming rotational broadening of 50 km \persec,
this works out to 0.6-3\tee{-17} \cgsflux \AA$^{-1}$.  This is
comparable to the flux observed in a variable single-peaked H$\alpha$
emission line, super-imposed on a double-peaked line -- present in
July, 1994, but absent in May, 1993 \cite{remillard96}.  However,
variability in RS CVn H$\alpha$ emission at this level has not been
reported.

Finally, we note that C {\sc IV} has been observed in the UV
spectra of RS~CVn sources, correlated with X-ray emission
\cite{dempsey93a}.  There are no reported C {\sc IV} lines of
SXTs in the literature, although it would seem that they should be
present.  

\subsection{Excess Optical/UV Light and ADAFs}
\label{sec:adaf} 

  Nearly all optical observations in quiescence find evidence for
excess optical/UV light over and above that from the companion (see 
\citenp{mcclintock99} for a recent example, and Table 2). 
The most commonly invoked source of this emission is an outer disk that is
truncated at some finite inner radius in order to avoid over producing
UV and X-ray emission. However, the advent of the ADAF models for the
quiescent X-ray emission has modified this story, as the synchrotron
emission from the inner parts of the ADAF flow can possibly explain
the observed optical excess  \cite{narayan97b}.  In
these models, the X-rays are made via Comptonization of the
synchrotron photons from deeper in the flow.  Indeed, an advantage of
the ADAF models (which can't {\it a priori} predict the X-ray flux) is
their ability to predict the ratio of the optical to the X-ray
emission in a nearly model independent way.  We begin by summarizing
the state of modeling for \v404cyg, as coronal emission appears
unlikely to explain its quiescent X-ray emission.

The current best set of ADAF models \cite{narayan97b,quataert99}
simultaneously fit the quiescent excess optical and X-ray emission
from \v404cyg. The required mass transfer rate in the quiescent
ADAF flow is comparable to the total mass transfer rate in the binary,
implying the incoming material from the companion is shared between
the hot advective flow and accumulation in the outer disk. All of the
excess optical/UV emission is explained as synchrotron emission from
the ADAF, and is emitted within a few tens of Schwarzschild radii of
the $\approx 10 M_\odot$ black hole.  The outer cool disk makes no
contribution to the total excess emission in this model.

  How do the ADAF models fare when a substantial part of the X-ray
emission comes from the corona of the companion?  The optical/UV
excess from A 0620-00 is well measured, and ADAF modeling that fit the
X-ray data over-predicted the optical/UV emission by factors as large
as four \cite{narayan97b}. This conflict is alleviated if some of the
observed X-rays are from the stellar corona, as then the ADAF
contribution can be reduced to explain the optical/UV excess. In that
case, the stellar corona and ADAF X-ray emission are comparable. 

An alternative picture would be to explain all the BHC X-ray emission as
coronal and the optical/UV excess as from a truncated outer disk. 
This  situation seems likely for GRO
J1655-40, as the current ADAF models that fit the X-ray data
under-predict the optical/UV excess \cite{hameury97}. This system
might best be explained with all X-ray
emission coming from the stellar corona.  Indeed, the detailed
modeling by Orosz \& Bailyn \cite*{orosz97} of the excess optical
emission exclusively in terms of an outer disk was successful and
required no ADAF flow. 

\section{Discussion and Conclusions}
\label{sec:conclude}

   We have shown that two (\nmon \ and \nsco) of the three X-ray
detected black hole binaries exhibit X-ray fluxes entirely consistent
with coronal emission from the companion star.  The upper limits on
the remaining BHCs are also consistent with production via
chromospheric activity in the secondary.  In addition, we found that
the photospheric lithium content measured for many of the SXT
companions is typical for coronally active stars.  We therefore
conclude that the quiescent X-ray emission from all but one BHC might
be due to coronal emission from the companion star.  There are many
ways to confirm the chromospheric hypothesis.  The most
straightforward is X-ray spectroscopy of the BHCs in quiescence with
\chandra and {\it XMM-Newton}. Resolving the soft (0.2-2.0 keV) X-ray
emission into the coronal line emission common for chromospheric
emission would be strong confirmation.  However, the calculated
equivalent widths make their detection quite difficult
\cite{narayan99}. 

  Additional means of confirming the chromospheric hypothesis is
analysis of the Ca {\sc II} H\&K absorption lines, the ``re-filling''
of which is correlated with the X-ray chromospheric surface flux
\cite{maggio87}; or from similar analysis of the Ca {\sc II} infra-red
triplet, as performed by Dempsey \etal\ \cite*{dempsey93c}.  Radio
emission (5-8 GHz) observed from coronally active stars is correlated
with X-ray luminosity; consistency between the $F_X/F_{\rm radio}$ of
BHCs and that observed from coronally active stars would support the
interpretation of the X-rays as due to stellar coronal activity.
Coronal variability from flares (which occur roughly 40\% of the time
in RS CVn systems; \citenp{osten99}) requires that such observations
be carried out simultaneously in the X-ray and radio frequencies.

  Not all quiescent SXT X-ray emission can be explained as coronal
activity in the companion. The quiescent X-ray flux from the BHC
\v404cyg\ is a factor of ten brighter than can be explained as coronal
emission and all four NSs (\aql, \cenx4, \xsix1608, \x2129) have
quiescent X-ray luminosities which are at least ten times greater than
expected from chromospheric emission alone.  As noted in the
introduction, much of the quiescent X-ray emission from the NS systems
is naturally explained as thermal emission from the NS
\cite{brown98,rutledge99,rutledge00}.

However, accretion power is the only choice for the unambiguous
black hole candidate \v404cyg.  The
current ADAF modeling requires accretion rates in quiescence of order
$\dot{M}\sim 10^{-9}-10^{-10} M_\odot \ {\rm yr^{-1}}$, a large
fraction of the mass transfer rate in the binary. Quiescent accretion
rates {\it at least three orders of magnitude lower} may be needed to
explain the observed variability in some of the quiescent NSs (a
factor of 4.2\ppm0.5 in 8 days from \cenx4; \citenp{campana97} and in
4U~2129+47, by a factor of 3.4\ppm0.6 between Nov-Dec 1992 and March
1994; \citenp{garcia99,rutledge00}). There is no simple 
reason why these accretion rates should be so different, although some
scenarios have been investigated \cite{menou99}. 

 Much of the current debate about the role of accretion as a power
source in quiescence centers around understanding what fraction of the
continuous flow into the outer disk (which makes itself apparent via
excess continuum emission and broad, double-peaked $H\alpha$ lines)
accumulates there (for the next outburst) versus proceeding all the
way into the compact object.  The ADAF modeling of the BHCs require
that the incoming flow is about evenly split, whereas the much more
severe limits on quiescent accretion for the NS systems says that
99.9\% of the accreted matter does not find its way to the compact
object. 

  The observational situation is clearly in a state of flux.  The
upcoming X-ray observations of the previously detected SXTs will sort
out the emission sources via X-ray spectral information, rather than
just total fluxes.  It might be that all SXT's have variable accretion
rates in quiescence and that the basal quiescent X-ray flux is set by
either coronal emission from the companion or -- when present -- by
thermal emission from the neutron star.

\acknowledgements

We thank Frank Verbunt for an encouraging discussion and for making us
aware of his early discussion of this idea. We thank Ramesh Narayan
and Feryal Ozel for discussions of ADAF modeling. We gratefully
acknowledge useful comments on this paper by Tom Ayres, Ed Brown, Phil
Charles, Mike Eracleous, Alex Filippenko, J.-P. Lasota, Jerry Orosz
and Frank Verbunt, and the anonymous referee. We thank Peter Wheatley
for making us aware of the results on V 471 Tauri.  L.B. was the CHEAF
Visiting Professor at the Astronomical Institute, ``Anton Pannekoek''
of the University of Amsterdam when this work was initiated and is a
Cottrell Scholar of the Research Corporation. This work was supported
in part by the National Science Foundation through Grant NSF94-0174
and NASA via grants NAGW-4517 and NAG5-3239.

\clearpage
\begin{deluxetable}{lccccccc}
\scriptsize
\tablewidth{500pt}
\tablecaption{\lxlv\ for the SXTs \label{tab:xrt}}
\tablehead{
\colhead{Object}  & 
\colhead{Sp. Type}	&
\colhead{$V_{\rm q}$}  &
\colhead{\av} & 
\colhead{B.C.\supp{1}}	&
\colhead{\fux\tablenotemark{b}} 	& 
\colhead{\porb} & 
\colhead{log($\frac{F_{\rm X, unabs.}}{F_{\rm bol}}$)+3.0} \\
\colhead{} &
\colhead{} &
\colhead{(mag)} &
\colhead{(mag)} &
\colhead{(mag)} &
\colhead{} &
\colhead{(d)}	&
\colhead{} 
}
\startdata
\multicolumn{8}{c}{Black Hole Candidates} \\
GRO J0422+32    &M2V\supp{2,15} & 22.8\supp{15} & 1.3\ppm0.2\supp{20}	& $-$1.8\ppm0.3   & $<$270& 0.211\supp{15} &	$<$4.5 \\   
A0620$-$00	&K4-K7V \supp{3}& 18.2\supp{16}	& 1.21\ppm0.06\supp{34}	& $-$0.3\ppm0.2	  & 2.4\ppm{1.4}& 0.323\supp{25} & 0.8\ppm0.3	\\
\nmus           &K3-K4V\supp{4} & 20.4\supp{17}	& 0.60\ppm0.16\supp{21} & $-0.35$\ppm0.2  & $<$1.7& 0.433\supp{26} & 	$<$2.2	\\    
4U 1543$-$47	&A2V\supp{5}	& 16.55\supp{5} & 1.55\ppm0.16\supp{5}	& $-$0.25\ppm0.2  & $<$16 & 1.12\supp{5}	 & 	$<$1.3   \\
GRO J1655$-$40	&F3-F6IV\supp{6}& 17.5\supp{35}	& 4.0\ppm0.3\supp{6}	& $-$0.04\ppm0.3  & 6\ppm1.2& 2.621\supp{27} &	$-$0.1\ppm0.3	\\
H 1705$-$25	&K3V\supp{7,31}	& 22\supp{7}	& 1.4\ppm0.5\supp{22}	& $-$0.33\ppm0.3  &$<$18 	&0.521\supp{7,31} &     $<$3.7 	\\ 
GS~2000+25      &K3-6V\supp{33,8}& 21.8\supp{18}& 4.65(\ppm0.3\tablenotemark{c})\supp{23}& $-$0.35\ppm0.3  &$<$0.57& 0.344\supp{32,33} & 	$<$0.7 	\\
\v404cyg        &K0IV\supp{9}	& 18.42\supp{19}& 4.0\ppm0.4 \supp{19}  & $-0.19\ppm0.1$  & 85\ppm20& 6.471\supp{28} & 1.4\ppm0.3	\\
\multicolumn{8}{c}{Neutron Stars} \\						   	  
Aql X-1         &K7V\supp{10}	& 21.6\supp{10}	& 1.5\ppm0.3\supp{10}	& $-$0.65\ppm0.3  & 88\ppm5& 0.789\supp{29} & 	3.5\ppm0.2 \\
Cen X-4         &K7V\supp{11,12,13}&18.5\supp{11,12}&0.3\ppm0.15\supp{11}& $-$0.65\ppm0.3 & 36\ppm4& 0.629\supp{12,13} & 	2.3\ppm0.2 \\
\xsix1608       &G0-K0 IV-V	&  (see text)	& (see text)		& $-0.13$\ppm0.3  & 160/1300\ud{40\%}{20\%}& --	&  2.8/3.7\ppm0.3  \\
\x2129          &F6IV \supp{14}	& 17.9\supp{14}	& 1.2\ppm0.16\supp{14} 	& $-$0.09\ppm0.3  & 12\ppm2/21\ppm7& 0.218\supp{30} &  1.5/1.7\ppm0.2 \\
\enddata
\tablerefs{
1, \citenp{zombeck}; 
2, \citenp{harlaftis99}; 
3, \citenp{murdin80};  
4, \citenp{casares97}; 
5, \citenp{orosz98}; 
6, \citenp{orosz97}; 
7, \citenp{remillard96}; 
8, \citenp{harlaftis96}; 
9, \citenp{wagner92}; 
10, \citenp{chev99}; 
11, \citenp{blair84}; 
12, \citenp{chev89}; 
13, \citenp{mcclintock90}; 
14, \citenp{deutsch96}; 
15, \citenp{filippenko95}; 
16, \citenp{mcclintock86}; 
17, \citenp{mcclintock92}; 
18, see Sec.~\ref{sec:2000}; 
19, \citenp{casares93};
20, \citenp{shrader92}; 
21, \citenp{gonzalez91}; 
22, \citenp{griffiths78}; 
23, \citenp{chev90}; 
25, \citenp{mcclintock86}; 
26, \citenp{rr92}; 
27, \citenp{bailyn95}; 
28, \citenp{casares94}; 
29, \citenp{chevalier98}; 
30, \citenp{thorstensen79}; 
31, \citenp{filippenko97}; 
32, \citenp{chev93}; 
33, \citenp{filippenko95b}; 
34, \citenp{wu76,oke77}; 
35,  \citenp{orosz95}
}
\tablenotetext{b}{$F_{X, unabs.}$ in \ee{-14} \cgsflux\  (0.4-2.4 keV)}
\tablenotetext{c}{Assumed value}

\end{deluxetable}

\begin{deluxetable}{lrrr}
\scriptsize
\tablewidth{400pt}
\tablecaption{Excess Continuum Flux Contributions ($f$) from Literature\label{tab:f}}
\tablehead{
\colhead{Ref.}  & 
\colhead{Obj.}  & 
\colhead{$f$\tablenotemark{a}(\%)}  & 
\colhead{Passband}  
}
\startdata
\citenp{casares93}	&\v404cyg	&$<$10	&$R$ band	\\
\citenp{casares94}	&\v404cyg 	&$<$13		&6400-6600\AA 	\\
\citenp{callanan96}	&GS 2000+25	&$<$12		&$K'$(19500-23500\AA)\\
\citenp{harlaftis96}	&GS 2000+25	&6\ppm5		&5700-6800\AA	\\
\citenp{filippenko97}	&N Oph 1977	&70-90\tablenotemark{c}	&6300\AA	\\
\citenp{harlaftis97}	&N Oph 1977	&63-74\tablenotemark{c}	&6250\AA	\\
\citenp{rr96}		&N Oph 1977	&40\ppm20	&5500\AA	\\
\citenp{orosz96}	&GS 1124-68	&39\tablenotemark{b}	&$B+V$ band	\\
			&GS 1124-68	&52\tablenotemark{b}	&$B+V$ band	\\
\citenp{casares97}	&GS 1124-68	&12-15	&$R$ band	\\
\citenp{oke77}		&A0620-00	&43\ppm6	&$V$ band	\\
\citenp{marsh94}	&A0620-00	&17\ppm4	&4400-5200\AA	\\
			&		&6\ppm4		&5800-6800\AA	\\
\citenp{mcclintock86}	&A0620-00	&40\ppm10	&5500\AA	\\
\citenp{mcclintock95}	&A0620-00	&56\ppm8\tablenotemark{b}&	5150-5300\AA\\
			&		&43\ppm4	&5150-5300\AA	\\
			&		&37\ppm5	&5150-5300\AA	\\
			&		&29\ppm12	&5150-5300\AA	\\
			&		&43\ppm7	&5150-5300\AA	\\
			&		&27\ppm5	&6350-6500\AA	\\
\citenp{filippenko95}	&GRO J0422+32	&30-60		&6300\AA	\\
\citenp{callanan96b}	&GRO J0422+32	&$<$30		& $V$ band	\\
\citenp{harlaftis99}	&GRO J0422+32	&39\ppm4	&5600-6900\AA	\\
\citenp{orosz98}	&4U 1543-47	&10\ppm5	&$B$ band	\\
			&		&21\ppm5	&$V$ band	\\
			&		&32\ppm5	&$R$ band	\\
			&		&39\ppm5	&$I$ band	\\
\citenp{orosz97}	&\nsco		&50\tablenotemark{b}	&$V$ band	\\
			&\nsco		&5\tablenotemark{b}	&$V$ band	\\
\citenp{filippenko99}	&N Vel 1993	&60-70		&6300\AA	\\
\citenp{chev89}		&Cen X-4	&$>$80		&3800\AA	\\
			&		&40-50		&$B$-band	\\
			&		&25-30		&$V$-band	\\
			&		&20		&6500\AA	\\
\citenp{mcclintock90}	&Cen X-4	&25-30		&$V$-band	\\
\enddata
\tablenotetext{a}{Fractional continuum flux contribution in listed passband}
\tablenotetext{b}{Observations taken on different days}
\tablenotetext{c}{Measurements from the same data}

\end{deluxetable}

\clearpage


\begin{figure}[htb]
 \caption{\label{fig:lxlbol} Flux ratio \lxlv\ vs. binary orbital
period for RS CVns (small points; Dempsey \etal 1993b), BHCs (filled
points), NSs (open points) and a best-fit relation found for rapidly
rotating, late-type isolated dwarfs (where we use their rotation
period; Singh \etal\ 1999).  Of the three detected BHCs, \nmon\ and
\nsco\ have \lxlv\ consistent with coronal emission
(\lxlv\approxlt\ee{-3}), while \v404cyg\ is at least an order of
magnitude above this. The four detected NSs are 1-3 orders of
magnitude above this limit as well.  }
\end{figure}
\nocite{singh99}

\begin{figure}[htb]
\caption{\label{fig:lars} X-ray flux vs. stellar bolometric flux for
X-ray detected BHCs (shaded regions) and 2$\sigma$ upper-limits. The
broken line is $F_x/F_{\rm bol}$=$10^{-3}$.  The bottom line is the
\chandra/ACIS-S 50ksec detection limits for a \nhtt=0.2, \ktr=1.0 keV
source (assuming 10 counts for a detection). Detection of X-rays from
GRO~J0422+32, GS~1124-68, and \hseventeen\ with \chandra\ would be
well above the level expected from coronal emission. \qzvul\ is close
to the \lxlv=$10^{-3}$ limit and may be detected.  While \ufifteen\ is
apparently well within X-ray detectable range, we expect none will be
detected (see text). }
\end{figure}

\begin{figure}[htb]
\caption{\label{fig:li} Distribution of detected lithium abundances in
chromospherically active systems, and the detected values for five
SXTs. The bin width is 0.5 dex in N(Li). The SXT abundances have
uncertainties of $\sim$0.5 in dex.  While the SXTs have, on average,
higher Li abundances by $\times10$, these are not such that Li
production mechanisms involving a compact object are required, as they
are well within the range observed from from RS CVns, which do not
have compact objects. }
\end{figure}

\clearpage
\pagestyle{empty}
\begin{figure}[htb]
\PSbox{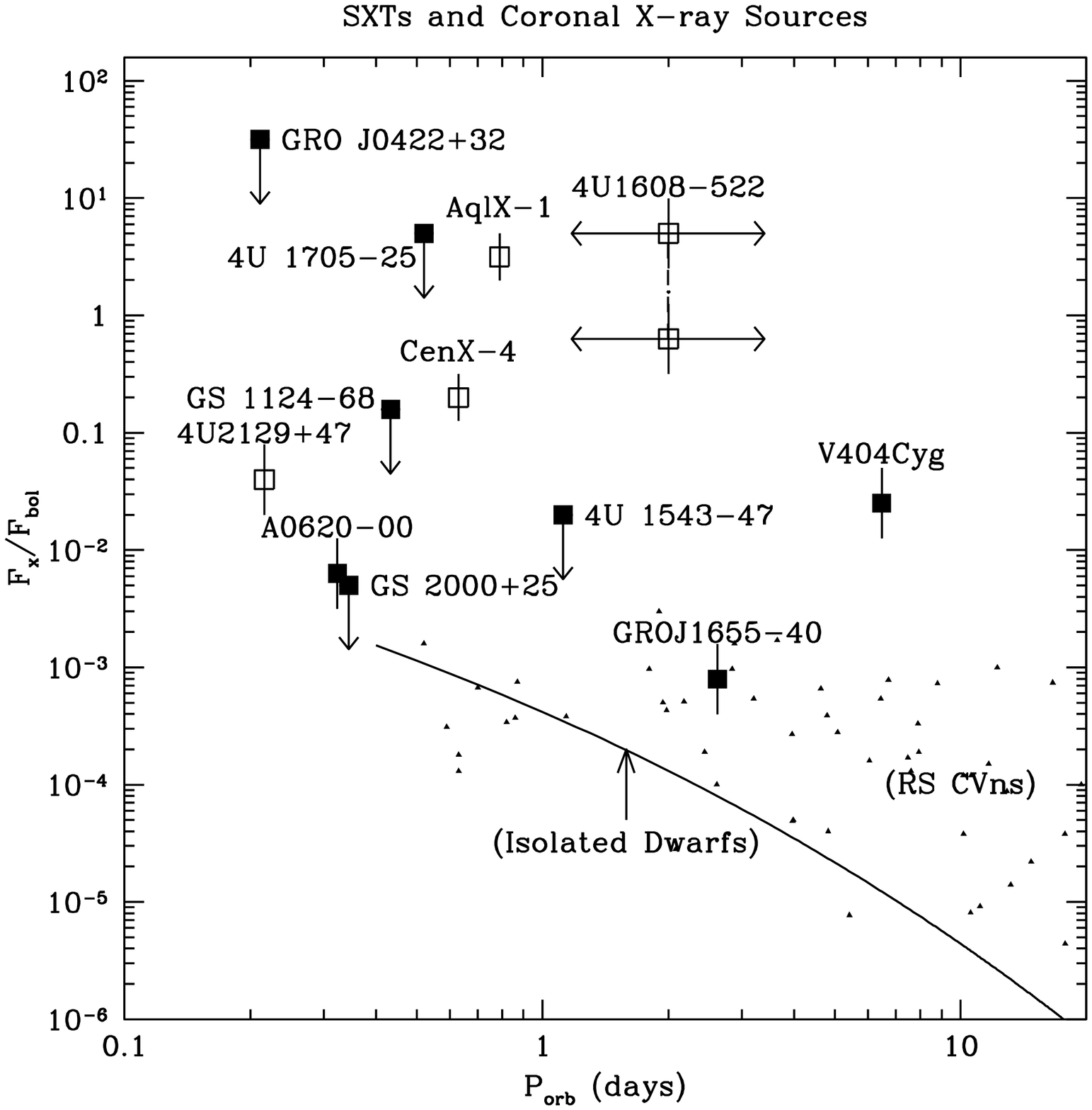 hoffset=-80 voffset=-80}{14.7cm}{21.5cm}
\FigNum{\ref{fig:lxlbol}}
\end{figure}

\clearpage
\pagestyle{empty}
\begin{figure}[htb]
\PSbox{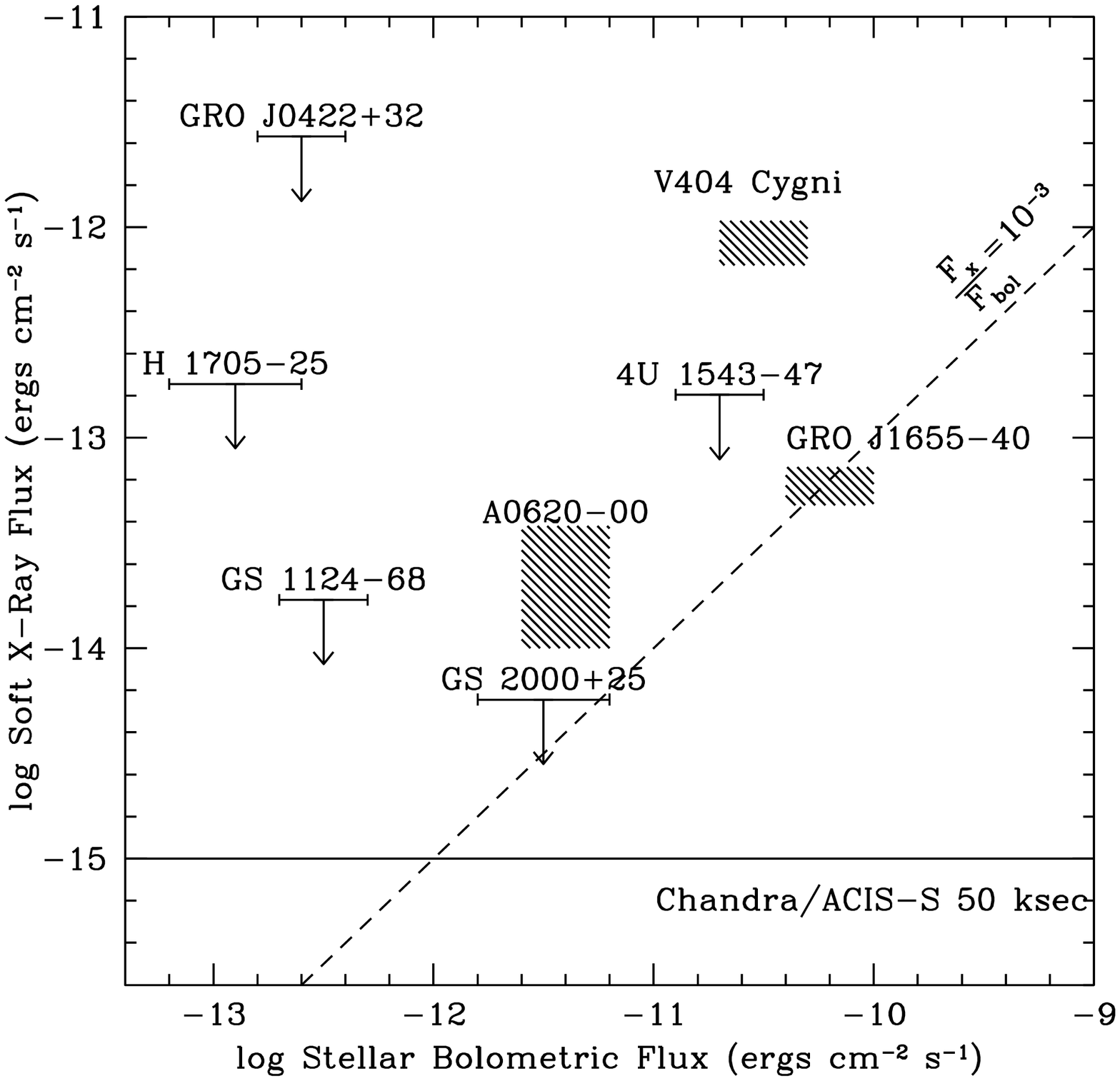 hoffset=-80 voffset=-80}{14.7cm}{21.5cm}
\FigNum{\ref{fig:lars}}
\end{figure}

\clearpage
\pagestyle{empty}
\begin{figure}[htb]
\PSbox{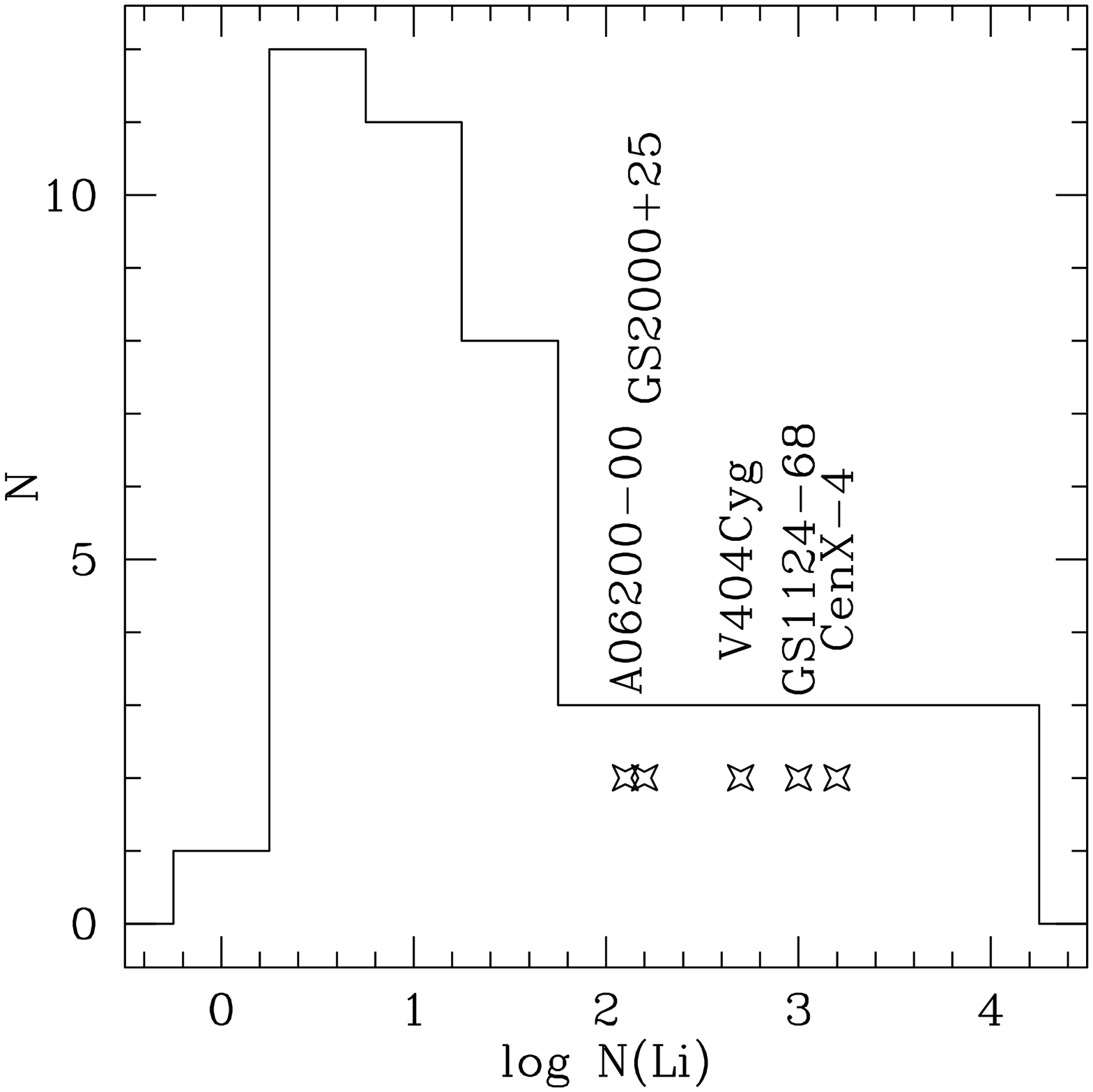 hoffset=-80 voffset=-80}{14.7cm}{21.5cm}
\FigNum{\ref{fig:li}}
\end{figure}

\end{document}